\documentstyle[aps,prb,epsf,graphicx]{revtex}
\begin{document}

\title{Formation of polaron clusters}
\author{C.A. Perroni$^{1,2}$, G. Iadonisi$^{1,2}$, and V.K. Mukhomorov$^{3}$}
\address{$^{1}$Coherentia-INFM, UdR di Napoli, via Cinthia 80126 Naples, Italy  \\
$^{2}$Dipartimento di Scienze Fisiche, Universit\`a "Federico II"
di Napoli, via Cinthia 80126 Naples, Italy \\
$^{3}$Agrophysical Institute, St. Petersburg 195220, Russia}

\maketitle

\begin {abstract}
The formation of spherical polaron clusters is studied within the
Fr$\ddot{o}$hlich polaron theory. In a dilute polaron
gas, using the non-local statistical approach and the polaron pair
interaction obtained within the Pekar strong coupling theory, the
homogeneous phase results to be unstable toward the appearance of polaron
clusters. The physical conditions of formation for the clusters
are determined calculating the critical values of electron-phonon
interaction for which bound states in the collective polaron
potential develop. Finally the sequence in the filling of the
states is found and the stability of the clusters is assessed.
\end {abstract}

\newpage

\section{Introduction}

In the last years the presence of strong electron-phonon coupling
and polaronic effects has been pointed out by many experimental
results in several compounds, such as high-temperature cuprate
superconductors, \cite{1new} colossal magnetoresistance
manganites, \cite{2new} nickelates and quasi-1D materials organic
conjugated polymers. \cite{3} Furthermore it has been debated if
the electron-phonon interaction can give rise in perovskite oxides
to charge-ordered states or to more complex electronic phases such
as stripes, strings or clustered states. \cite{5}

The large amount of experimental data has renewed the interest in
studying simplified electron-phonon coupled systems of the
Holstein \cite{4new} or Fr$\ddot{o}$hlich \cite{froh} type. The
formation of the polaron have been long studied in the frame of
the Fr$\ddot{o}$hlich model, where the polar long-range
interaction between electronic and ionic charges is taken into
account and the medium is considered continuous. For its relation
to theories regarding cuprate superconductors, large attention has
been devoted to the formation of the bipolaron (bound state of two
polarons), establishing the range of the values of the
electron-phonon coupling constant and of the dielectric parameters
of medium which allow its existence. \cite{19} Also the binding
energy, the effective mass, its internal structure and optical
features have been studied. \cite{14,15,16,17,18} Actually many
methods have been used to study the bipolaron formation in the
Fr$\ddot{o}$hlich scheme: Lee-Low-Pines approach for intermediate
values of the electron-phonon coupling constant, \cite{15} Pekar
polaron strong coupling theory, \cite{14,17,18} and path-integral
technique. \cite{16} They give the possibility to define an
effective polaron-polaron potential, which is repulsive both at
large and small distance between the particles, while it is
attractive at intermediate distance. This polaron-polaron
interaction has been used as starting point for many-body
calculations finding that, for densities smaller than those
typical of metals, the many-polaron system exhibits a charge
density instability in the intermediate coupling regime.
\cite{giulio}

A related research activity concerns the study of the conditions
under which electronic or polaronic Wigner crystals,
\cite{11,11bis} polaron molecules or clusters \cite{kusma1} and
strings \cite{kusma2} can form. Indeed in the regime of very low
densities the Wigner crystal of electrons can transform into a
polaronic crystal by increasing the electron-phonon coupling.
\cite{11} However increasing the density at strong electron-phonon
coupling the Wigner crystal of polarons becomes unstable.
\cite{11bis} Such an instability suggests that novel types of
electronic or polaronic structures such as clusters, molecules or strings
can be stabilized by a strong electron-phonon interaction. Actually the
molecules can arise at conditions determined by the
structure of an effective interaction at short and long distances.
\cite{kusma1} Their behavior is very quantum, and they can assume
a stringed form. It has been suggested that in perovskite oxides,
such as manganites and cuprates, these molecules  could represent the
intermediate step of more complex structures such as
stripes. \cite{kusma2} Finally the electronic and ionic structure
of Metal-Ammonia solutions \cite{9,10} represents a system where
polaronic clusters can be actually observed.

In this work we start from the knowledge of the effective
polaron-polaron interaction within the Pekar strong coupling
theory. \cite{14,15,17,21} Using the non-local statistical
approach, \cite{18,20} we show that in the homogeneous phase the
mode due to the collective polaron potential is unstable
suggesting that polaron clusters can form in the system. Next we
show that there is  a range of values of the electron-phonon
interaction and of the static and high frequency dielectric
constants which allow the formation of clusters with a number of
polarons larger than two. Finally we discuss some features of the
cluster such as its radius and its stability.

In sect. II we discuss the basic equations; in sect. III we
analyze the collective excitations of the homogeneous polaron phase and its
instability  due to the attractive interaction between polarons;
in sect. IV we indicate the conditions for which the polaron
cluster forms, we discuss the shell structure in the
cluster and finally we study the stability of the polaron
clusters.

\section{Basic equations}

In the system we have two types of particles, the polarons and
the ions, with distribution functions $f_{p}({\vec r},{\vec v},t)$
and $f_{i}({\vec r},{\vec v},t)$, respectively, in the
single-particle phase space. In the non-local and non-linear
Vlasov kinetic equations \cite{21} the distribution functions
$f_{p}({\vec r},{\vec v},t)$ and $f_{i}({\vec r},{\vec v},t)$ are
solutions of the equations

\begin{equation}
\frac{\partial f_{p}({\vec r},{\vec v},t) }{\partial t}+{\vec v }
\cdot {\vec \nabla_r} f_{p}({\vec r},{\vec v},t) -\frac{1}{m_p}
{\vec \nabla_r}\left[ U({\vec r},t)-e \Phi({\vec r},t)  \right]
\cdot {\vec \nabla_v} f_{p}({\vec r},{\vec v},t)  = 0  \label{e1}
\end{equation}

\begin{equation}
\frac{\partial f_{i}({\vec r},{\vec v},t)}{\partial t}+{\vec v }
\cdot {\vec \nabla_r} f_{i}({\vec r},{\vec v},t) -\frac{e}{m_i}
{\vec \nabla_r} \Phi({\vec r},t)  \cdot {\vec \nabla_v}
f_{i}({\vec r},{\vec v},t) = 0, \label{e2}
\end{equation}
where $m_p$  and  $m_i$ are the effective masses of
polaron and ion, respectively, $e$ is the electron charge, and
$U({\vec r},t)$ is the collective self-consistent potential
\begin{equation}
U({\vec r},t)=\int d{\vec r}^{\prime} d{\vec v}^{\prime} K(|{\vec
r}-{\vec r}^{\prime}|) f_{p}({\vec r}^{\prime},{\vec
v}^{\prime},t), \label{e3}
\end{equation}
with the kernel $K$ indicating the polaron-polaron potential
depending only on the relative distance between particles. The
distribution function $f_p$ is determined not only by the polaron
collective potential $U$ but also by the electrostatic potential
$\Phi$ arising from the non-punctual compensation between the
polaron and ionic charges. Indeed the potential $\Phi$ satisfies
the Poisson equation
\begin{equation}
\Delta \Phi({\vec r},t)=-\frac{4 \pi e}{\epsilon_s} \rho({\vec
r},t), \label{e4}
\end{equation}
where the density $\rho({\vec r},t)$ is defined as
\begin{equation}
\rho({\vec r},t)=\rho_i({\vec r},t)-\rho_p({\vec r},t)=\int d{\vec
v} f_{i}({\vec r},{\vec v},t)- \int d{\vec v} f_{p}({\vec r},{\vec
v},t). \label{e5}
\end{equation}
We note that the force acting on the ions is related only to the
electrostatic potential $\Phi$ screened by the static dielectric
constant $\epsilon_s$.

In order to solve the Vlasov equations, the polaron-polaron potential $K$ has
to be specified. We use the potential $K$ obtained within the Pekar
strong coupling bipolaron theory \cite{14,17,18,22} which has been
adapted in order to find reliable results in the intermediate to
strong coupling regime. This polaron-polaron potential is
attractive at intermediate distances, this behaviour resulting by
the Fr$\ddot{o}$hlich interaction and effects of quantum
mechanical exchange and inter-electron correlations. Moreover the potential
depends on the ratio $\epsilon^*/\epsilon_{\infty}$, where $\epsilon^{*}$ is
defined through the equation
$1/\epsilon^{*}=1/\epsilon_{\infty}-1/\epsilon_{s}$,  with
$\epsilon_{\infty}$ high frequency dielectric constant. In Fig. 1
we show the potential $K$ for different values of the ratio
$\epsilon^*/\epsilon_{\infty}$ when the bipolaron is in its
singlet ground-state and the asymptotic energy of the two free
polarons has been subtracted. Throughout the paper, the values
$1.10$, $1.08$, $1.05$, $1.02$ and $1.00$ are used for the ratio
$\epsilon^*/\epsilon_{\infty}$, i.e. the ratio
$\epsilon_s/\epsilon_{\infty}$ varies from $10$ to infinity.

The potential $K$ obtained within the Pekar strong coupling theory
can be approximated by the following analytical expression
\begin{equation}
K(r)=2 \alpha^2 \hbar \omega_0 \left[
\frac{(\epsilon^*/\epsilon_{\infty}-1) a^{*}_{0}}{r}+K_1(r)e^{-
\delta r} \right], \label{e6}
\end{equation}
where $K_{1}(r)$ is given by
\begin{equation}
K_{1}(r)=d+c r^2+[a+b(r-\rho_0)^2]\left(1-e^{-\gamma r}
\right)-\frac{(\epsilon^*/\epsilon_{\infty}-1) a^{*}_{0}}{r}.
\label{e7}
\end{equation}
In Eq. (\ref{e6}) $\alpha=(1/2 \epsilon^*)(e^2/\hbar
\omega_0)(2m^*\omega_0/\hbar)^{1/2}$ is the dimensionless
electron-phonon coupling constant, $m^*$ is the effective mass at
the bottom of the conduction band, $\omega_0$ is the longitudinal
optical phonon frequency in the long wave-length approximation,
and $a^{*}_{0}=\alpha^{-1}(\hbar/2 m^* \omega_0)^{1/2}$ is the
effective Bohr radius. The parameters $a$, $b$, $c$, $d$,
$\gamma$, $\delta$, and $\rho_0$ of the potentials in Eq.
(\ref{e6}) and  (\ref{e7}) have been obtained by an accurate fit
of the polaron-polaron potential in the Pekar theory.
\cite{14,17,18,22}and are listed in Table \ref{tab1}. In the
numerical calculations it is furthermore assumed
$\epsilon_{\infty}=2$, $m^* = m$ with $m$ electron mass, and
$\hbar \omega_0 =0.03 eV$.

The Vlasov equations in Eqs. (\ref{e1}) and (\ref{e2}) are valid
in the classical limit and the use of polaron-polaron interaction
(\ref{e6}) is accurate for a dilute system. Therefore the
temperature $T$ has to be high compared with the degeneracy
 temperature of polarons:  $T> \hbar^2 N_p^{2/3}/m_p k_B$,
with $N_p$ polaron density and $k_B$ Boltzmann constant. This
condition is verified at reasonable temperatures if $N_p < 5
\times 10^{18} cm^{-3}$. We note that for these densities some
energy scales, for example the polaron plasmon frequency, are smaller
or of the same order than the optical phonon frequency. Clearly,
for the temperatures considered above, quantum effects in the
statistics can be neglected and the Boltzmann-Maxwell distribution
function can be used at the thermodynamic equilibrium.

\section{Collective excitations of the polaron system}
In this section we analyze the collective excitations of the
polaron system. As discussed in the following subsection $A$, the polaron
plasmon is obtained neglecting the role of the collective
potential $U$ in the Vlasov equations (\ref{e1}) and (\ref{e2}).
On the contrary in subsection $B$ we will examine the collective
excitations induced by the potential $U$ in absence of the
electrostatic potential $\Phi$. These density oscillations are
completely different from the polaron plasmons, since at small
values of the momentum $k$ they have a dispersion relation
approximately proportional to $k$.

The procedure of calculation of these two types of excitations is
the same. Indeed only the role of the polarons is mainly
considered and the Vlasov equations are linearized taking
\begin{equation}
f_{p}({\vec r},{\vec v},t)=f_0({\vec v})+\phi({\vec r},{\vec
v},t), \label{e8}
\end{equation}
where $f_0({\vec v})$ is the Maxwell distribution function
\begin{equation}
f_0({\vec v})=N_p \left( \frac{m_p}{2 \pi k_B T } \right)^{3/2}
e^{-\frac{m_p v^2}{2 k_B T}}
 \label{e9}
\end{equation}
associated to the homogeneous spatial distribution of the
particles and $\phi({\vec r},{\vec v},t)$ is strongly smaller than
$f_0({\vec v})$. Therefore it is assumed that the perturbations
induced by the time-dependent potentials are weak.

\subsection{Polaron plasmons}
The polaron plasmons are the collective excitations in the limit
of collective potential $U$ zero. In this case the external force
affecting the polaron distribution can be simply related to an
electrical field ${\vec E}({\vec r},t)$ defined as ${\vec E}({\vec
r},t)=-{\vec \nabla} \Phi({\vec r},t)$. Neglecting second order
terms, the equation (\ref{e1}) becomes
\begin{equation}
\frac{\partial \phi({\vec r},{\vec v},t) }{\partial t}+{\vec v }
\cdot {\vec \nabla_r} \phi({\vec r},{\vec v},t)= \frac{e}{m_p}
{\vec E}({\vec r},t) \cdot {\vec \nabla_v} f_{0}({\vec v}).
\label{e10}
\end{equation}
Fourier transforming in space and time, one obtains
\begin{equation}
\phi({\vec k},{\vec v},\omega)= \frac{e {\vec E}({\vec
k},\omega)}{i m_p ({\vec k} \cdot {\vec v} -\omega)}  \cdot {\vec
\nabla_v} f_{0}({\vec v}), \label{e11}
\end{equation}
where the coefficient of ${\vec \nabla_v} f_{0}({\vec v})$
represents the amplitude of momentum that the polaron acquires
under the field ${\vec E}$. Clearly this quantity has to be small
compared with the mean momentum obtained through the
equilibrium distribution $f_0({\vec v})$.

Using Eq. (\ref{e5}), the Fourier transform of particle density
$\rho$ can be evaluated. This last quantity is connected to the
polarization ${\vec P}$ of the system since $i {\vec k} \cdot
{\vec P}=-\frac{e}{\epsilon_s} \rho$. Employing the pole-rule
defined by Landau, \cite{landau} we obtain the relation
\begin{equation}
i {\vec k} \cdot {\vec P}({\vec k},\omega)=-\frac{e^2}{\epsilon_s}
{\vec E}({\vec k },\omega) \cdot \int d{\vec v} \frac{{\vec
\nabla_v} f_{0}({\vec v})}{i m_p ({\vec k} \cdot {\vec v}
-\omega)-i \eta}, \label{e12}
\end{equation}
where $\eta$ is infinitesimal. If the field ${\vec E}$ and ${\vec
P}$ are directed along ${\vec k}$, then the longitudinal
dielectric constant $\epsilon_l({\vec k}, \omega)$ can be derived
by means of the equation $4 \pi {\vec P} = (\epsilon_l -1 ){\vec
E} $, yielding
\begin{equation}
\epsilon_l({\vec k}, \omega)=1-\frac{4 \pi e^2}{\epsilon_s k^2}
\int d{\vec v} \frac{{\vec k} \cdot {\vec \nabla_v} f_{0}({\vec
v})}{i m_p ({\vec k} \cdot {\vec v} -\omega)-i \eta}. \label{e13}
\end{equation}
The dielectric constant of Eq. (\ref{e13}) is a complex quantity
implying that the energy of the electrical field can be dissipated
in the medium (Landau damping). Hence longitudinal electrical
waves can propagate through the system and their dispersion
relation is obtained by the equation $\epsilon_l({\vec k},
\omega)=0$. Supposing $\omega \gg k v_T$, with $v_T=(k_B
T/m_p)^{1/2}$ mean quadratic velocity along the direction of
propagation, the zeros of the equation can be derived. At low
order in momentum the real part of the frequency is
\begin{equation}
\omega=\omega_{pp}(1+\frac{3}{2}k^2 a_{p}^{2}),
\end{equation}
with $a_p=(\epsilon_s k_B T/4 \pi N_p e^2)^{1/2}$ and
$\omega_{pp}$ polaron plasma frequency defined as
\begin{equation}
\omega_{pp}=v_T/a_p=\sqrt{\frac{4 \pi N_p e^2}{\epsilon_s m_p}},
\label{e14}
\end{equation}
while the imaginary part is exponentially small. As expected in
the limit $\omega_{pp} \leq \omega_0$, the plasma frequency is
screened by the static dielectric constant $\epsilon_s$.
\cite{mahan}

In this derivation we have taken into account only the polaron
contribution. Within the same approach it would be
possible to consider also the role of the ionic plasmons through
Eq. (\ref{e2}) and the interplay between polaronic and ionic
oscillations.  \cite{landau}
However, it is more interesting to focus on the
collective excitations due to the polaron collective potential
$U$.

\subsection{Excitations due to the collective potential $U$}

In this subsection we consider the time-dependent kinetic
equations (\ref{e1}) for the polaron distribution function
neglecting the contribution from the electrostatic potential $\Phi$. We
investigate the propagation of longitudinal waves with lengths
larger than the polaron-polaron distance in the bipolaron. Therefore our
aim will be the calculation of the dispersion relationship
linking frequency $\omega$ and wave vector $k$ in the limit $k
\rightarrow 0$.

If one neglects second order terms, the Vlasov equation (\ref{e1})
becomes
\begin{equation}
\frac{\partial \phi({\vec r},{\vec v},t) }{\partial t}+{\vec v }
\cdot {\vec \nabla_r} \phi({\vec r},{\vec v},t)= \frac{1}{m_p}
{\vec \nabla_v} f_{0}({\vec v}) \cdot {\vec \nabla_r} \int d{\vec
r}^{\prime} d{\vec v}^{\prime} K(|{\vec r}-{\vec r}^{\prime}|)
\phi({\vec r}^{\prime},{\vec v}^{\prime},t). \label{e15}
\end{equation}
The solution of the integro-differential equation (\ref{e15}) can
be written in the following form \cite{21}
\begin{equation}
\phi({\vec r},{\vec v},t)=\frac{1}{2 m_p} {\vec \nabla_v}
f_{0}({\vec v}) \cdot {\vec \nabla_r} \int d{\vec r}^{\prime}
\left[ \int_{t_0}^{t} d \tau K(|{\vec r}-{\vec v}(t-\tau)-{\vec
r}^{\prime}|) \rho({\vec r}^{\prime},\tau)+ \int_{t_1}^{t} d \tau
K(|{\vec r}-{\vec v}(t-\tau)-{\vec r}^{\prime}|) \rho({\vec
r}^{\prime},\tau) \right], \label{e16}
\end{equation}
with $t_0 \leq t \leq t_1$ and $\rho({\vec r},t)$ defined in Eq.
(\ref{e5}). Integrating with respect to the velocity, one obtains
\begin{equation}
\rho({\vec r},t)=\frac{1}{2 m_p} \int d {\vec v}  {\vec \nabla_v}
f_{0}({\vec v}) \cdot {\vec \nabla_r} \int d{\vec r}^{\prime}
\left[ \int_{t_0}^{t} d \tau K(|{\vec r}-{\vec v}(t-\tau)-{\vec
r}^{\prime}|) \rho({\vec r}^{\prime},\tau)+ \int_{t_1}^{t} d \tau
K(|{\vec r}-{\vec v}(t-\tau)-{\vec r}^{\prime}|) \rho({\vec
r}^{\prime},\tau) \right]. \label{e17}
\end{equation}
Considering the spatial and temporal Fourier transform of the
density $\rho$, Eq. (\ref{e17}) becomes
\begin{equation}
\rho({\vec k},\omega)= \frac{1}{2} \int_{- \infty}^{t} d \tau
e^{-i \omega (t-\tau)} G ({\vec k},t-\tau) \rho({\vec k},\omega)+
\frac{1}{2} \int_{t}^{\infty} d \tau e^{-i \omega (t-\tau)} G
({\vec k},t-\tau) \rho({\vec k},\omega), \label{e18}
\end{equation}
where $G ({\vec k},t)$ is defined by
\begin{equation}
G ({\vec k},t)=\frac{i \sigma(k)}{m_p} {\vec k} \cdot \int d {\vec
v} e^{i {\vec k} \cdot {\vec v} t}{\vec \nabla_v} f_{0}({\vec v}),
\end{equation}
with $\sigma(k)$ Fourier transform of the polaron-polaron
potential shown in Fig. 2 for different values of the ratio
$\epsilon^*/\epsilon_{\infty}$. The condition of existence of the
nontrivial solutions of the equation (\ref{e18}) is the following
\begin{equation}
\int_{0}^{\infty} d t G ({\vec k},t) \cos(\omega t) =1,
\end{equation}
that links implicitly the frequency $\omega$ to the momentum
${\vec k}$. We obtain the dispersion for small $k$ in
the form
\begin{equation}
\omega=\left( \frac{v_T}{2^{1/6}} \right) k \left(1 + \frac{k_B
T}{N_p \sigma(k)} \right)^{1/3}, \label{e19}
\end{equation}
where $v_T=(k_B T/m_p)^{1/2}$ is the mean quadratic velocity along
the direction of propagation. Therefore the system exhibits a collective mode
sustained by the coherent self-consistent interaction arising from neighboring
particles.
Clearly these excitations are well defined if
\begin{equation}
1 + \frac{k_B T}{N_p \sigma(k)}>0. \label{e40}
\end{equation}
As deduced from Fig. 2, the collective mode can certainly
propagate for values of $k$ around 0 where $\sigma(k)$ is
positive. The Fourier transform $\sigma$ at $k=0$ is given by the
spatial integral of the potential $K$ that, as shown in Fig. 1,
becomes completely negative as the ratio
$\epsilon^*/\epsilon_{\infty}$ approaches the unity. By fixing the
ratio $\epsilon^*/\epsilon_{\infty}$ and increasing the
electron-phonon coupling constant $\alpha$, the potential
$\sigma(k)$ deepens implying that the region of stability in $k$
is reduced. So there is a range of values of the momentum where
Eq. (\ref{e40}) is violated. This suggests the possibility that
the particles in equilibrium are not longer single polarons, but,
due to the attractive effect of the interaction, a polaron of the
system can bind another one in order to form a bipolaron. Clearly
two particles could form with a third a cluster of more than two
particles. Therefore it is important to study directly the
formation of polaron clusters induced by the collective potential
$U$.

\section{Polaron clusters}
In this section we will study the solutions of the Vlasov
equations (\ref{e1}) and (\ref{e2}) in the temporal stationary
regime. Clearly these equations are satisfied if the polaron $f_p$
and ion $f_i$ distribution functions are independent of position.
Our aim is to study the solutions of equations (\ref{e1}) and
(\ref{e2}) near to the homogeneous ones.

In stationary conditions, we write for the polaron and ion
distribution function
\begin{equation}
f_j ({\vec r},{\vec v})=\rho_j ({\vec r}) w_j ({\vec v}), \label{e20}
\end{equation}
where $j$ stands for $p$ or $i$, $\rho_j ({\vec r})$ and $w_j
({\vec v})$ the spatial and velocity distribution functions,
respectively. If the velocity distribution function is that of the
equilibrium, then from equations (\ref{e1}) and (\ref{e2}) the
spatial distribution functions of polarons and ions can be deduced as
\begin{equation}
\rho_p ({\vec r})=
N_p
e^{- \frac{U({\vec r})-e \Phi({\vec r})}{K_B T} }
\label{e22}
\end{equation}
and
\begin{equation}
\rho_i ({\vec r})=N_i
e^{- \frac{e \Phi({\vec r})}{K_B T} }, \label{e23}
\end{equation}
where $N_i$ is the ion concentration, such that $N_p=N_i=N_0$ since the system
has charge neutrality.
Substituting (\ref{e22}) and (\ref{e23}) in (\ref{e3}) and (\ref{e4}), we
obtain the coupled equations
\begin{equation}
U({\vec r})= \int d{\vec r}^{\prime} K(|{\vec r}-{\vec
r}^{\prime}|)  \rho_{p}({\vec r}^{\prime})
\label{e24}
\end{equation}
and
\begin{equation}
\Delta \Phi({\vec r})=\left( -\frac{4 \pi e }{\epsilon_s} \right)
\left[\rho_{p}({\vec r})-\rho_{i}({\vec r})\right], \label{e25}
\end{equation}
that allow to determine self-consistently both the collective polaron
$U({\vec r})$ and electrostatic $\Phi({\vec r})$  potentials.

The exact solution of Eqs. (\ref{e24}) and (\ref{e25}) is a
formidable task. Therefore, considering the dependence of the
potential $K$ on the relative distance, we iteratively build a
solution for a spherical symmetric cluster of radius $R$ obtaining
at the lowest order $\Phi_0(r)=0$ and
\begin{equation}
 U_0(r)= \left\{ \begin{array}{ll}
 N_0 \int d{\vec r}^{\prime} K(|{\vec r}^{\prime}-{\vec r}|)|,  & r \leq R   \\
 N_0 \int d{\vec r}^{\prime} K(|{\vec r}^{\prime}-{\vec r}|)|,  &
 r>R.
\end{array} \right. \label{e26}
\end{equation}
We note that the collective potential $U_0(r)$ is directly
proportional to the particle density $N_0$ and through $K$ it
depends on the electron-phonon coupling constant $\alpha$. Through
the use of the coordinate transformation
\begin{equation}
s^2=r^2+x^2-2 x r \cos(\theta),  s ds = x r \sin(\theta) d \theta,
\end{equation}
the potential inside the sphere $U_0^{(in)}(r)$ is written as
\begin{equation}
U_0^{(in)}(r)=\frac{2 \pi N_0}{r}
\left[ \int_{0}^{r} dx x \int_{r-x}^{r+x} ds s K(s) +
\int_{r}^{R} dx x \int_{r-x}^{r+x} ds s K(s) \right], r \leq R ,
\end{equation}
and connects with continuity to that outside  $U^{(out)}(r)$
\begin{equation}
U_0^{(out)}(r)=\frac{2 \pi N_0}{r}
\int_{0}^{R} dx x \int_{r-x}^{r+x} ds s K(s), r > R.
\end{equation}
The local modifications of the spatial distribution function due
to the attractive forces are described by the polaron spatial
density $N_1 (r)$
\begin{equation}
N_1(r)=N_0 \left(- \frac{U_0(r)}{k_B T}\right)  \label{e27}
\end{equation}
showing that the distribution of the polarons is determined by the
collective potential $U_0(r)$. Actually the charge neutrality is
locally perturbed and the increase of the polaron density can be
consistent with a bound cluster of particles.

In Figs. 3(a) and 3(b) we show the collective potential for different
radii R and for the value of $\epsilon^*/\epsilon_{\infty} =1.05 $.
The qualitative behaviours are similar
in all the considered cases except for $\epsilon^*/\epsilon_{\infty}=1.00$.
For a fixed ratio $\epsilon^*/\epsilon_{\infty}$, $U_0(r)$
is negative in a region around $r=0$, it assumes the minimum value at $r=0$
and becomes positive at large values of $r$. The quantity $U_0(0)$ is first
decreasing and then increasing as function of $R$, so that it exists
a value of ${\bar R}$ which gives the minimum value of $U_0(0)$. Only in the case
$\epsilon^*/\epsilon_{\infty}=1$ (i.e. $\epsilon_s=\infty$)
$U_0(0)$ is always decreasing as function of R.
In Fig. 4 we show the quantity $U_0(0)$ as function of $R$ for all the
indicated values of $\epsilon^*/\epsilon_{\infty}$, except for
$\epsilon^*/\epsilon_{\infty}=1$. It occurs that the decrease of
$\epsilon^*/\epsilon_{\infty}$ implies the increase of ${\bar R}$.
We give to ${\bar R}$  the meaning of the best dimension of the
cluster.

It is important to calculate the minimum values of $\alpha$ for
which the cluster begins to form. We have evaluated the critical
values $\alpha_{c}$ within the full quantum mechanical approach of
a particle in the collective potential $U_0(r)$ calculating when
the first bound level develops. \cite{30} Hence the quantity
$\alpha_c$ indicates the minimum value for which the bipolaron
forms. It has been checked that these values are only slightly
smaller than those obtained in the next subsection through the
statistical model that can be generalized at arbitrary levels. In
particular we will investigate the resulting shell structure in
the cluster.

\subsection{Shell structure of the polaron cluster}

In analogy with the structure of nuclei \cite{27} and metal
clusters, \cite{28} we try now to calculate the conditions for
existence of the polaron clusters and the related shell structure.
For the spherically  symmetric collective potential $U_0(r)$, we
use the same criterion of the atomic physics regarding the quantum
numbers to introduce in order to characterize the state of a
single electron in the atoms. In other words we introduce the
quantum numbers of single particle states without considering
neither the electrostatic corrections nor the relativistic ones.
In the statistical model, \cite{27,29} the shells with the orbital
quantum number $l$ ($l=0, 1, 2, 3...$) begin to form when $N_l$
defined by
\begin{equation}
N_l=\frac{4}{\pi \hbar} \left(l+\frac{1}{2} \right) \int_{R_1}^{R} d r
\left[ -2 m_p U_0(r) -\frac{\hbar^2 (l+1/2)^2}{r^2}\right]^{1/2}=
\frac{4}{\pi \hbar} \left(l+\frac{1}{2} \right) \int_{R_1}^{R} d r
\left[ P^2 -\frac{\hbar^2 (l+1/2)^2}{r^2}\right]^{1/2}
\label{e28}
\end{equation}
becomes larger than $1$. In the above formula $R$ is the cluster
radius, $R_1$ is the lower value of $r$ for which the function to
integrate is zero, $P$ is the maximum polaron momentum. In the
statistical approximation the maximum momentum $P$ is connected
with the local polaron density $N_1 (r)$ given in (\ref{e27})
through the relation
\begin{equation}
P(r)=2 \pi \hbar \left( \frac{3 N_1(r)}{8 \pi} \right)^{1/3},
\label{e30}
\end{equation}
so that, writing $N_1 (r) = m \rho_0 (r)$, where $m$ is the number
of the particles involved in the cluster, we have to ensure that
$\int d {\vec r} \rho_0 (r)= 1$. Taking into account Eq. (\ref{e30}), we find
the following equation
\begin{equation}
\frac{1}{r}=-\frac{1}{3} \frac{d}{dr} \ln \rho_0 (r),
\end{equation}
which allows to determine explicitly the value $R_1$.

In table \ref{tab2} we have reported the calculated values of the
electron - phonon coupling constants $\alpha_{c,l}$ from which the shell with
the indicated value of $l$ begins to be filled. We find that, increasing the
polaron concentration, the collective potential $U_0(r)$ deepens, so that the
values of $\alpha_{c,l}$ decrease. For example, taking
$\epsilon^*/\epsilon_{\infty}=1.05$, for
$N_0 = 10^{18} cm^{-3}$ the first shell $l= 0$ starts to be filled as soon as
$\alpha_{c,0}=10.2$, while for $N_0 = 10^{17} cm^{-3}$ we have
$\alpha_{c,0}=18.2$. Furthermore, since the increase of the number $l$
reflects a higher binding energy of the particles in the cluster, in table
\ref{tab2} the values of  $\alpha_{c,l}$ get enhanced as a function of $l$.

Finally we can calculate the sequence of the shell formation in a
polaron cluster. In analogy to the filling of states for
nuclear matter and metal clusters,\cite{27,28} we obtain
the sequence $(1s)^2 (2p)^6 (3d)^{10} (2s)^2 (4f)^{14} $ for the
polaron cluster. This order differs from the known order of
filling of electron levels in atoms: in fact shells with 2, 8, 18,
20 polarons are stable. Then it is possible that in the system stable
clusters can form becoming the basic units in equilibrium at a given
temperature.

Finally we can make a rough estimate of $T_{cr}$, the highest
temperature under which the clusters can form or, better, the
lowest temperature above which the polaron phase is homogeneous.
Actually, knowing now the critical electron-phonon coupling
constants that trigger the cluster formation, we can calculate the
temperatures for which the collective mode discussed in the
previous section becomes unstable signaling the possibility of
polaron clusters. The stability condition of the collective mode
of the homogeneous phase is violated when $\sigma(k)$ vanishes
becoming negative at larger values of $k$. Therefore the polarons
are the carriers at equilibrium approximately for temperatures
\begin{equation}
T> T_{cr}=\frac{N_0 |\sigma(k)|_{min}}{k_B}, \label{e31}
\end{equation}
where $|\sigma(k)|_{min}$ is the modulus of the Fourier transform
of the polaron-polaron potential $K$ calculated for the value of
$k$ corresponding to its minimum value (the attractive part drives
the instability) and for the critical values $\alpha_{c,l}$ of the
coupling constant.  In the table \ref{tab3} the critical
temperatures $T_{cr}$ are shown. We note that the critical
temperature obtained in Eq. (\ref{e31}) is a function of the
polaron density also through the constants $\alpha_{c,l}$ that, as
reported in tab. \ref{tab2}, are strongly dependent on $N_0$.
Since the collective potential responsible for the cluster
formation is proportional to $N_0$, we expect that the critical
temperatures strongly increase as a function of the polaron
concentration, as confirmed by the results of tab. \ref{tab3}.
Clearly it is possible to define the critical temperatures for
condensation of polarons for different shells. Increasing the
quantum number $l$, the binding energies becomes larger, the
cluster becomes more stable, so that the critical temperatures are
enhanced.

In this section we have realized that, at the lowest order of the
iterative procedure used for solving equations (\ref{e24}) and
(\ref{e25}), the electrostatic potential $\Phi$ defined in Eq.
(\ref{e4}) does not affect the cluster formation. However, at
higher orders, the difference between the density of polarons and
ions could perturb the states of the cluster. The frequency scale
of such perturbation is close to the frequency $\omega_{pp}$ of
the polaron plasma oscillation given in Eq. (\ref{e14}). Such
fluctuations can contribute to the broadening of the states
determined in the cluster. Obviously the broadening is small if
$\omega_{pp} < \omega_l$, where $\omega_l$ is the frequency of
motion of the particle in the collective potential at the level
with a quantum number $l$. From this inequality it is possible to
find the highest value of  the electron - phonon coupling constant
${\bar \alpha}$, such that for $\alpha < {\bar \alpha}$ the
cluster is unstable against these perturbations. For example, at
the concentrations $N_0=10^{17} cm^{-3}$ and $N_0=10^{18} cm^{-3}$
for $l=0$ and $\epsilon^*/\epsilon_{\infty}=1,08$, we have  ${\bar
\alpha}=16.6$ and ${\bar \alpha}=12.5$, respectively. These
results do not contradict the data listed in the table \ref{tab2}.
Therefore when the polaron cluster is formed in the system, it is
also stable against the energy fluctuations of the bound states of
the collective potential $U_0$.

\section{Conclusion}

Using the statistical model, we have found that clusters can form
in a dilute polaron gas. We have  calculated the critical values
of the electron-phonon interaction for which the ground and
excited state shells can be filled. The found sequence of the
shells is analogous to that of the nuclei and of the metal
clusters rather than that of the electron states in atoms. We have
also calculated the stability of the polaron clusters against the
fluctuations of the collective potential. We have found that the
values for which the cluster can form are sufficiently large to
assure the stability of the system. Finally the critical
temperature for the condensation of the polarons in the cluster
has been calculated. Our calculation scheme based on the strong
coupling Fr$\ddot{o}$hlich electron-phonon theory is coherent with
the results since all critical values for the formation of the
clusters are found in the regime of large coupling. Clearly this
implies that the theoretical results can be discussed in
connection with real systems only in situations of poorly screened
strong electron-phonon coupling. \cite{9,10}

The system with polaron clusters is characterized by collective
modes different from those of the homogeneous phase, for example
that induced by perturbations which tend to deform the cluster. In
such a case the difficulty is to solve the time dependent
equations starting at the zero order from the spatially perturbed
polaron distribution function calculated in the previous section.
Actually vibronic modes have been discussed in electronic bound
configurations showing that their frequency is of the order
$10^{14}-10^{15} Hz$ for typical parameters of metallic oxides.
\cite{kusma1} Finally we note that, in the quantum case of high
density and low temperature characteristic of perovskite oxides,
the system can exhibit a true phase transition toward a clustered
state that in the one-dimensional case can be a string. \cite{kusma2}

\section*{Figure captions}
\begin {description}

\item{Fig.1}
Polaron-polaron potential $K$ (in units of $2 \alpha^2 \hbar
\omega_0$) as a function of the distance $\rho$ (in units of
$a^{*}_{0}$) for different values of the ratio
$\epsilon^*/\epsilon_{\infty}$: solid
($\epsilon^*/\epsilon_{\infty} = 1.00$), large dot
($\epsilon^*/\epsilon_{\infty} = 1.02$), dash
($\epsilon^*/\epsilon_{\infty} = 1.05$), dash-dot
($\epsilon^*/\epsilon_{\infty} = 1.08$), small dot
($\epsilon^*/\epsilon_{\infty} = 1.10$).

\item{Fig.2}
Fourier transform $\sigma$ (in units of $2 \alpha^2 \hbar
\omega_0$ $a^{* 3}_{0}$) of the polaron-polaron potential $K$ as a
function of the momentum $k$ (in units of $a^{* -1}_{0}$)  for
different values of the ratio $\epsilon^*/\epsilon_{\infty}$:
1-solid ($\epsilon^*/\epsilon_{\infty} = 1.10$), 2-dot
($\epsilon^*/\epsilon_{\infty} = 1.08$), 3-dash-dot
($\epsilon^*/\epsilon_{\infty} = 1.05$), 4-dash
($\epsilon^*/\epsilon_{\infty} = 1.02$), 5-solid
($\epsilon^*/\epsilon_{\infty} = 1.00$).

\item{Fig.3}
(a) The collective polaron potential $U_0$ (in units of $2 \alpha^2
\hbar \omega_0 /N_0$) as a function of the distance $r$ (in units of
$a^{*}_{0}$) for $\epsilon^*/\epsilon_{\infty}=1.05$:  1-solid
(radius of the cluster $R=7 a^{*}_{0}$), 2-dash  (radius of the
cluster $R=11 a^{*}_{0}$), 3-dash-dot (radius of the cluster $R=15
a^{*}_{0}$). Optimal radius of the cluster is $15 a^{*}_{0}$. The
dot curve defines the boundaries of the cluster.

(b) The collective polaron potential $U_0$ (in units of $2 \alpha^2
\hbar \omega_0 /N_0$) as a function of the distance $r$ (in units of
$a^{*}_{0}$) for $\epsilon^*/\epsilon_{\infty}=1.05$:  1-solid
(radius of the cluster $R=16 a^{*}_{0}$), 2-dash  (radius of the
cluster $R=20 a^{*}_{0}$), 3-dash-dot (radius of the cluster $R=22
a^{*}_{0}$). Optimal radius of the cluster is $15 a^{*}_{0}$. The
dot curve defines the boundaries of the cluster.

\item{Fig.4}
The collective polaron potential $U_0$ (in units of $2 \alpha^2
\hbar \omega_0/N_0$) at $r=0$ as a function of the cluster radius $R$
(in units of $a^{*}_{0}$) for different values of the ratio
$\epsilon^*/\epsilon_{\infty}$: 1-dot
($\epsilon^*/\epsilon_{\infty} = 1.10$), 2-dash-dot
($\epsilon^*/\epsilon_{\infty} = 1.08$), 3-dash
($\epsilon^*/\epsilon_{\infty} = 1.05$), 4-solid
($\epsilon^*/\epsilon_{\infty} = 1.02$).

\end {description}

\newpage

\begin{table}
\caption{Parameters of the polaron-polaron potential $K$}
\centering
\begin{tabular}{|c|c|c|c|c|c|c|c|}
\hline   $\epsilon^*/\epsilon_{\infty}$  &  $a$ & $b$ ($a^{*
-2}_{0}$) & $c$ ($a^{* -2}_{0}$) & $d$ & $\gamma$ ($a^{* -1}_{0}$)
& $\delta$ ($a^{* -1}_{0}$) & $\rho_{0}$ ($a^{*}_{0}$)
\\ \hline
1.10    & $-7.95 \times 10^{-7}$ & $6.875 \times 10^{-3}$ &
$-6.821 \times 10^{-3}$ &  $1.4 \times 10^{-3}$   &  0.2475    &
0.2047 & $7.95 \times 10^{-2}$ \\ \hline 1.08  & $-5.175 \times
10^{-7}$  & $6.885 \times 10^{-3}$ & $-6.857 \times 10^{-3}$ &
0.001 & 0.245  & 0.20 & $8.17 \times 10^{-2}$ \\ \hline 1.05 & $
-1.907 \times 10^{-5} $  &   $ 6.855 \times 10^{-3} $ & $ -6.852
\times 10^{-3}$  &  $ 8.583 \times 10^{-4}$ & 0.25 & 0.19 & $ 8.81
\times 10^{-2}$ \\ \hline 1.02 & $ -6.0 \times 10^{-5}$  & $
6.9188 \times 10^{-3}$ &  $-6.887 \times 10^{-3}$  & $ -3.097
\times 10^{-3} $ &  0.286 & 0.166 & $9.25 \times 10^{-2}$ \\
\hline 1.00 & $ -7.92 \times 10^{-4} $ & $6.709 \times 10^{-3}$ &
$-6.677 \times 10^{-3}$ &  $ -7.750 \times 10^{-2}$ & 0.32 &  0.15
& 0.1087 \\ \hline
\end{tabular}
\label{tab1}
\end{table}

\begin{table}
\caption{The critical values  of the electron-phonon coupling
constant when the shell  with momentum $l$ is filled } \centering
\begin{tabular}{|c|c|c|c|c|}
\hline  $\epsilon^*/\epsilon_{\infty}$ &
\multicolumn{4}{c |}{Electron-phonon coupling constant $\alpha_{c,l}$} \\
\hline $N_0 = 10^{16} cm^{-3}$ & $l=0$ & $l=1$ & $l=2$ & $l=3$ \\
\hline 1.10 & 76.0 & 89.8 &  110.7 & 129.4  \\ \hline 1.08 & 51.5
& 59.8 & 73.7 & 86.1
\\ \hline 1.05 & 32.2 & 35.0 & 42.2  & 48.9 \\ \hline 1.02 & 23.7
& 25.4 & 30.4 & 35.1 \\ \hline 1.00 & 17.5 & 18.3 & 21.6 & 24.7 \\
\hline $N_0 = 10^{17} cm^{-3}$
& \multicolumn{4}{c |}{} \\
\hline 1.10 & 42.7 & 50.5 & 62.3 & 72.7 \\ \hline 1.08 & 29.0 &
33.6 & 41.5 & 48.4 \\ \hline 1.05 & 18.2 & 19.7 & 23.7 & 27.5 \\
\hline 1.02 & 13.3 & 14.3 & 17.1 & 19.7 \\ \hline  1.00 & 9.8 &
10.3 & 12.2 & 13.9 \\ \hline $N_0 = 10^{18} cm^{-3}$ &
\multicolumn{4}{c |}{} \\  \hline 1.10 & 24.0 & 28.4 & 35.0 & 40.9
\\ \hline 1.08 & 16.3 & 18.9 & 23.3 & 27.2 \\ \hline 1.05 & 10.2 & 11.1 & 13.4 &
15.5 \\ \hline 1.02 & 7.5 & 8.0 & 9.6 & 11.1 \\ \hline 1.00 & 5.6
& 5.8 & 6.8 & 7.8 \\ \hline
\end{tabular}
\label{tab2}
\end{table}

\begin{table}
\caption{Critical temperatures (in units of $2 \alpha^2 \hbar
\omega_0 /k_B$) corresponding to the filling of the cluster shell
with momentum $l$.} \centering
\begin{tabular}{|c|c|c|c|c|}
\hline $\epsilon^*/\epsilon_{\infty}$ &
\multicolumn{4}{c |}{Critical temperature $T_{cr}$} \\
\hline  $N_0 = 10^{16} cm^{-3}$ & $l=0$ & $l=1$ & $l=2$ & $l=3$ \\
\hline 1.10 & 2.4 & 3.1 & 4.7 & 6.5  \\ \hline 1.08 & 1.6 & 2.0 &
3.0 & 4.1
\\ \hline 1.05 & 0.91 & 1.1 & 1.6 & 2.1
\\ \hline 1.02 & 0.72 & 0.83 & 0.6 & 0.8 \\ \hline 1.00 & 0.84 & 0.92 & 1.2 & 1.7
\\ \hline $N_0 = 10^{17} cm^{-3}$
& \multicolumn{4}{c |}{} \\
\hline 1.10 & 7.0 & 9.8 & 15.0 & 28.4 \\ \hline 1.08 & 4.9 & 6.2 & 9.5 & 12.9
\\ \hline 1.05 & 2.9 & 3.4 & 4.9 & 6.7 \\ \hline 1.02 & 2.3 & 2.6 & 3.8 & 5.0
\\ \hline  1.00 & 2.6 & 2.9 & 4.1 & 5.3
\\ \hline $N_0 = 10^{18} cm^{-3}$ & \multicolumn{4}{c |}{}
\\  \hline 1.10 & 22.2 & 31.1 & 47.2 & 64.4
\\ \hline 1.08 & 15.6 & 19.6 & 29.8 & 40.6 \\ \hline 1.05 & 9.2 & 10.8 & 15.8 & 21.1
\\ \hline 1.02 & 7.2 & 8.2 & 11.8 & 15.8 \\ \hline 1.00 & 8.6 & 9.3 & 12.7 & 16.7
 \\ \hline
\end{tabular}
\label{tab3}
\end{table}

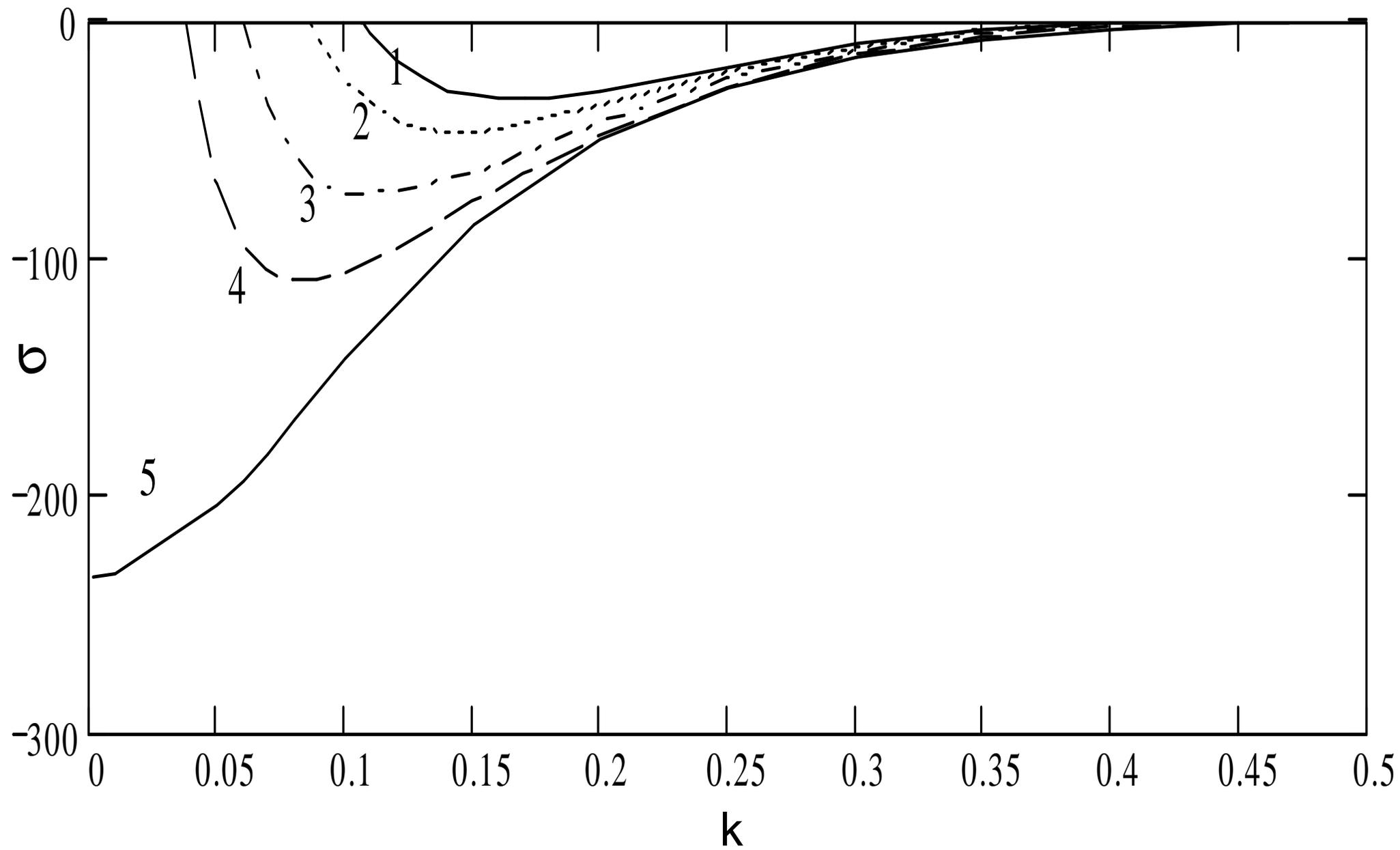
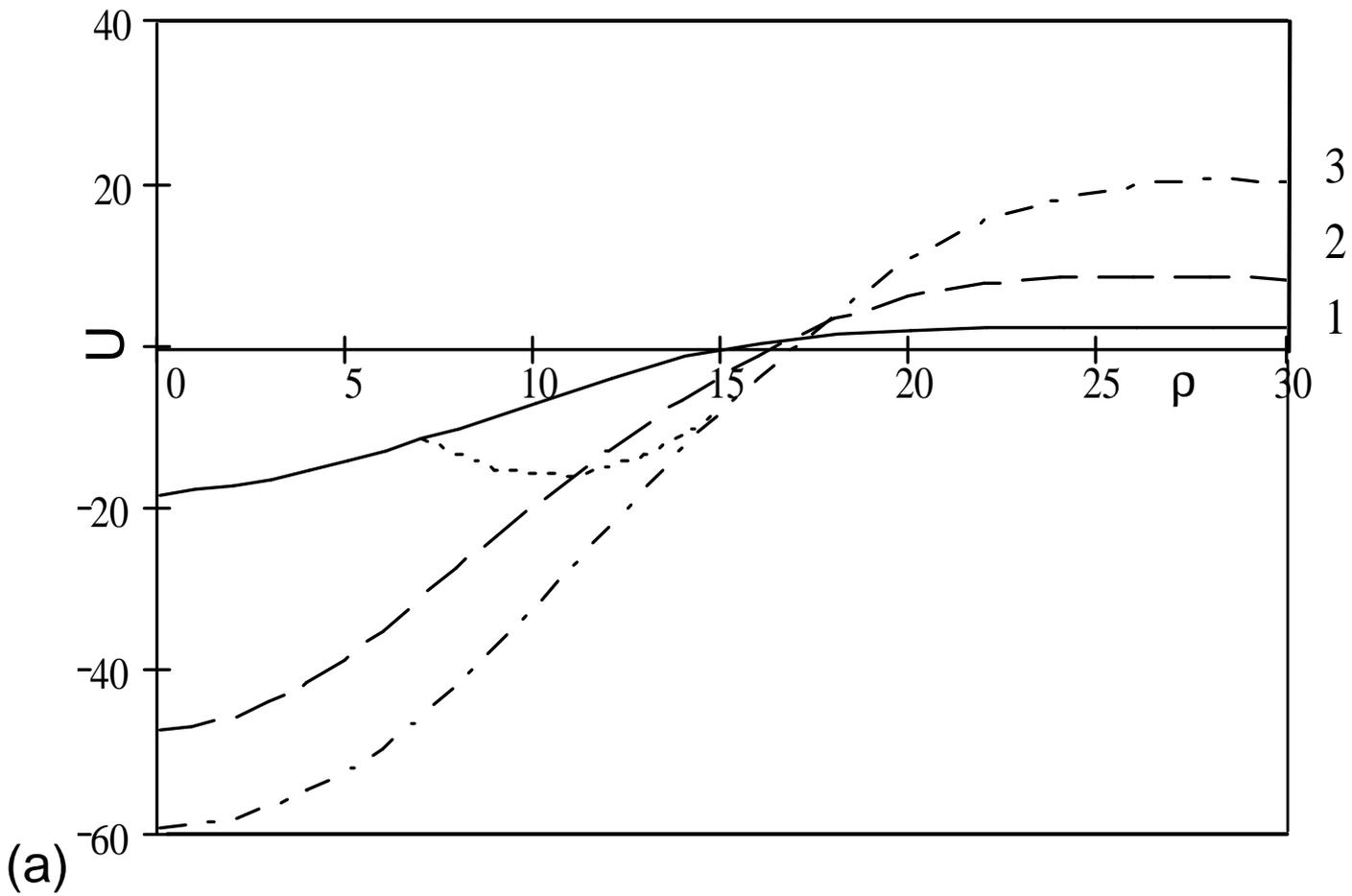
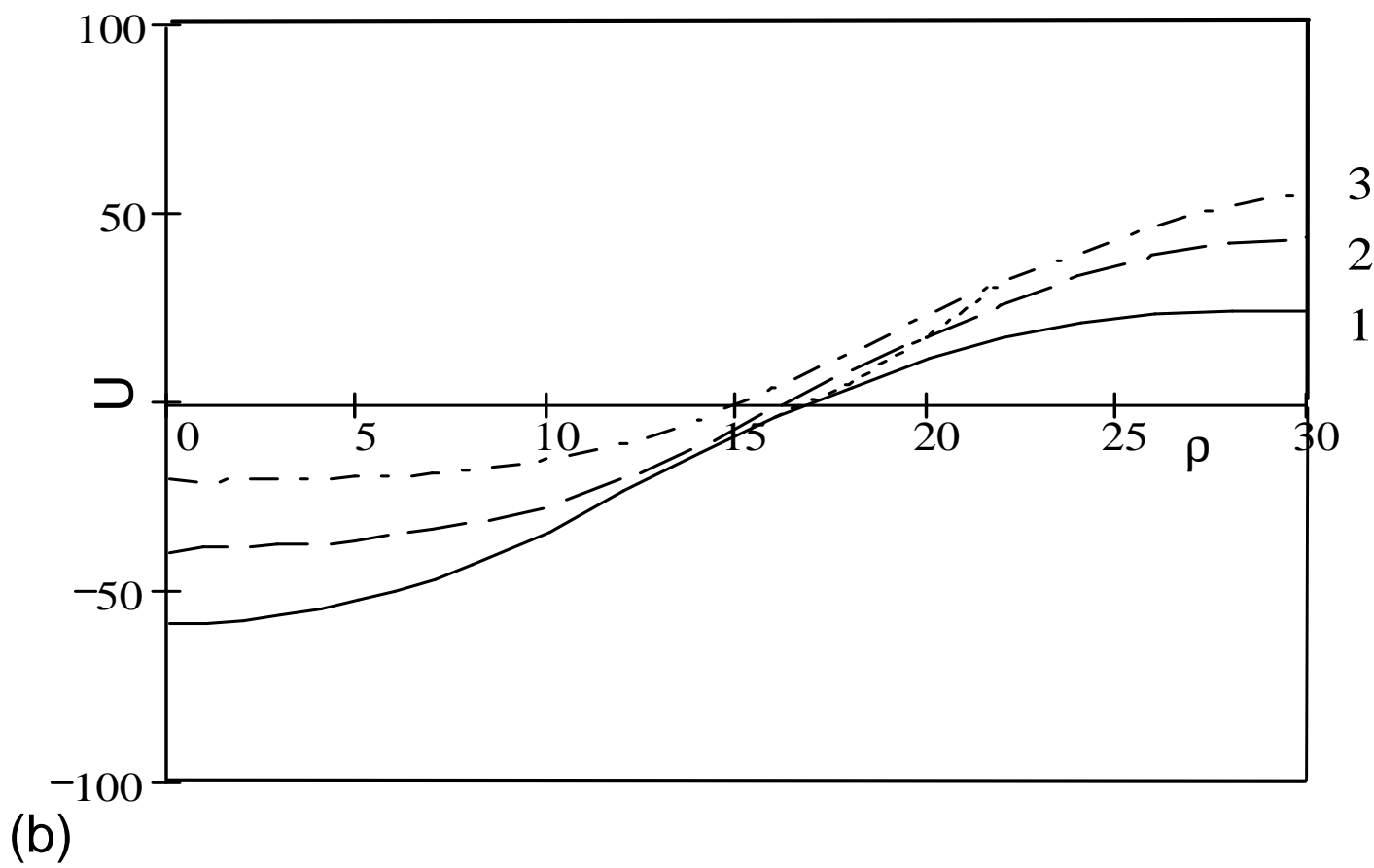
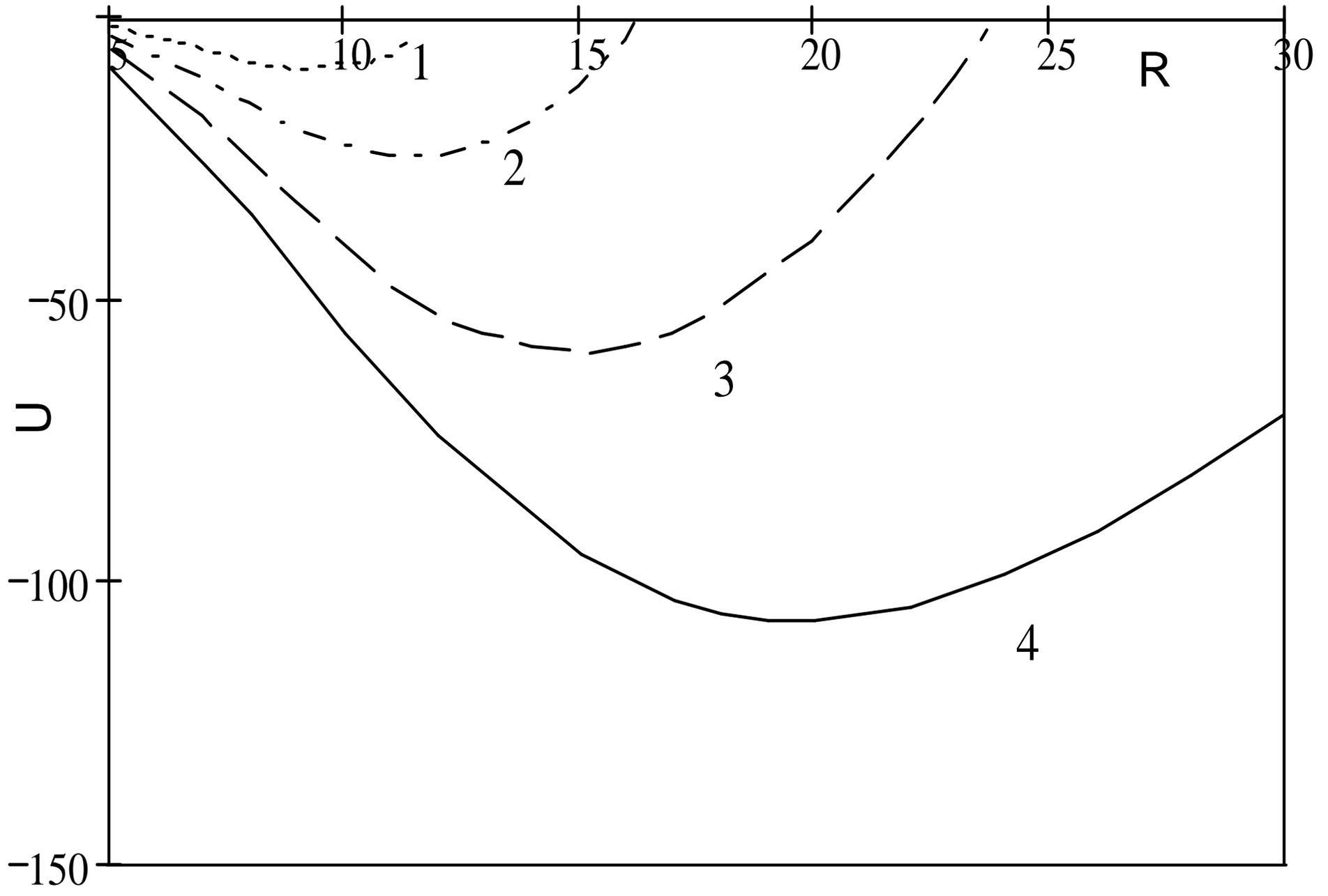
\end{document}